\documentclass[twocolumn,showpacs,preprintnumbers,amsmath,amssymb]{revtex4}
\usepackage{dcolumn}
\usepackage{bm}
\usepackage{graphicx,subfigure,xcolor}

\usepackage{braket}
\usepackage{xcolor}

\usepackage{color, soul}
\setstcolor{red}

\newcommand{\wo}{\omega_0}
\newcommand{\wk}{\omega_k}
\newcommand{\wkp}{\omega_{k'}}
\newcommand{\skj}{\sum_{\textbf{k}j}}
\newcommand{\skjkj}{\sum_{\textbf{k}\textbf{k'}jj'}}
\newcommand{\akj}{a_{\textbf{k}j}}
\newcommand{\akpjp}{a_{\textbf{k'}j'}}
\newcommand{\adkj}{a^{\dag}_{\textbf{k}j}}
\newcommand{\fkj}{\textbf{f}_{\textbf{k}j}}
\newcommand{\fkpjp}{\textbf{f}_{\textbf{k'}j'}}
\newcommand{\ekj}{\hat{e}_{\textbf{k}j}}

\begin{document}

\title{Dynamical Casimir-Polder force between an excited atom and a conducting wall}
\author{Federico Armata$^{1}$\footnote{f.armata@imperial.ac.uk}, Ruggero Vasile$^{2}$\footnote{ruggero.vasile@ambrosys.de}, Pablo Barcellona$^{3}$\footnote{pablo.barcellona@physik.uni-freiburg.de},
Stefan Yoshi Buhmann$^{3,4}$\footnote{stefan.buhmann@physik.uni-freiburg.de},\\
Lucia Rizzuto$^{5,6}$\footnote{lucia.rizzuto@unipa.it}, and Roberto Passante$^{5,6}$\footnote{roberto.passante@unipa.it}}
\address{$^1$QOLS, Blackett Laboratory, Imperial College London, London SW7 2BW, United Kingdom}
\address{$^2$UP Transfer GmbH an der Universit\"{a}t Potsdam, Am Neuen Palais 10, 14469 Potsdam, Deutschland}
\address{$^3$Physikalisches Institut, Albert-Ludwigs-Universit\"at Freiburg, Hermann-Herder-Str. 3, 79104 Freiburg,
Germany}
\address{$^4$Freiburg Institute for Advanced Studies, Albert-Ludwigs-Universit\"at Freiburg, Albertstr. 19, 79104 Freiburg, Germany}
\address{$^5$Dipartimento di Fisica e Chimica, Universit\`{a} degli Studi di Palermo and CNISM, Via Archirafi 36, I-90123 Palermo, Italy}
\address{$^6$INFN, Laboratori Nazionali del Sud, I-95123 Catania, Italy}


\begin{abstract}
We consider the dynamical atom-surface Casimir-Polder force in the non-equilibrium configuration of an atom near a perfectly conducting wall, initially prepared in an excited state with the field in
its vacuum state.
We evaluate the time-dependent Casimir-Polder force on the atom, and find that it shows an oscillatory behavior from attractive to repulsive both in time and in space. We also investigate the
asymptotic behavior in time of the dynamical force and of related local field quantities, showing that the static value of the force, as obtained by a time-independent approach, is recovered for
times much larger than the timescale of the atomic self-dressing, but smaller than the atomic decay time. We then discuss the evolution of global quantities such as atomic and field
energies, and their asymptotic behavior. We also compare our results for the dynamical force on the excited atom with analogous results recently obtained for an initially bare ground-state atom. We
show that new relevant features are obtained in the case of an initially excited atom, for example much larger values of the dynamical force with respect to the static one, allowing for an easier way
to single-out and observe the dynamical Casimir-Polder effect.
\end{abstract}

\maketitle

\section{\label{sec:level1}Introduction}
One of the most surprising consequences of quantum electrodynamics is the emergence of long-range interactions between neutral macroscopic objects and/or atoms/molecules, originating from quantum
zero-point vacuum fluctuations or exchange of field quanta \cite{Casimir48,CP48,Buhmann-book}.
In fact, the presence of boundaries, given for example by neutral macroscopic bodies and/or atoms, modifies the vacuum field fluctuations. This results in a measurable force between them, known as
Casimir and Casimir-Polder force. Casimir and Casimir-Polder forces have stimulated intense theoretical and experimental investigations since their discovery, and many theoretical models have been
proposed in the literature, highlighting fundamental properties such as their dependence on the shape and magnetodielectric properties of the interacting objects \cite{BKMM09,BW07} or
magnetic effects \cite{Haakh2009}. Recent experiments have confirmed with remarkable accuracy the theoretical predictions under a broad range of conditions such as temperature, geometry and
dielectric properties of the objects \cite{Lamoreaux05}, although some fundamental open questions pose theoretical and experimental challenges \cite{Milton04,KMM09}. Nowadays, it is recognized that
Casimir and Casimir-Polder effects play an important role in many different areas of physics, ranging from atomic physics \cite{Buhmann-book} and molecular biophysics \cite{PPT15}, to cosmology
\cite{Martin12}.
Being dominant at nanoscale separations, Casimir forces have also a significant role in the interaction with nanostructured materials \cite{Bender14} and in applications in micro- and
nano-technologies, for example microelectromechanical devices \cite{CAKBC01}.

New effects appear when a boundary, such as a reflecting or dielectric plate, is set in motion with a nonuniform acceleration: in this case a dynamical Casimir effect is set up and pairs of real
photons are produced in a sort of parametric excitation of the vacuum \cite{Moore70,Dodonov10}. A dynamical Casimir-Polder effect resulting from the optomechanical coupling of an oscillating
reflecting plate and a dilute gas of Rydberg atoms has been also recently proposed \cite{ABCNPRRS14}.
Even if an analogous of the dynamical Casimir effect has been observed in the context of superconducting circuits \cite{Wilson11} and Bose-Einstein condensates \cite{Jaskula12}, the experimental
observation of the dynamical Casimir effect due to an oscillating macroscopic boundary is still a challenge, because very high oscillation frequencies are necessary to obtain a measurable number of
real photons.
Alternative schemes have been recently proposed, where the mechanical motion of the boundary is replaced by a suitable modulation of the optical properties of a cavity boundary \cite{Agnesi09}.

New features occur in the dynamical Casimir-Polder effect between two atoms \cite{RPP04} or between an atom and a conducting mirror \cite{VP08,MVP10}.
A dynamical Casimir-Polder force may originate from many different nonequilibrium physical conditions, for example a dynamical self-dressing due to a nonequilibrium initial state, a
popolation-induced dynamics related to atomic spontaneous decay, motion of the atoms or of the macroscopic bodies or even a time-dependent change of the matter-radiation coupling due, for example, to
changes of dielectric properties of a macroscopic body.  All these situations are basically different from the dynamical Casimir effect, because the creation of real photons from the vacuum is not
directly involved. In the case of dynamical self-dressing,
the system evolves from a non-equilibrium quantum state such as a bare or partially dressed ground state of the atom, and the atom-wall Casimir-Polder force has a temporal evolution with new
peculiarities with respect to the static case, such as temporal oscillations, including possibility of transient repulsive forces. Such a scenario occurs also when some parameter involved in the
system Hamiltonian changes instantaneously, for example a sudden change of the atomic transition frequency induced by Stark shift through an external electric field suddenly switched on or off. In
this case, the state of the system before the change is no longer an eigenstate of the new Hamiltonian, and thus a time-evolution of the Casimir-Polder force is obtained \cite{MVP10}. Very recently
these studies have been extended to the cases of a real surface, where the excitation of surface plasmons plays an important role in the dynamical interaction \cite{HHSRP14} or to the case of an atom
in a cavity with a dielectric medium \cite{YZZSP14} or a chiral molecule near a chiral plate \cite{BPRB16}. All these studies show that, in the dynamical case, the Casimir-Polder force can be much
stronger compared with the static case around the round-trip time, that is the time taken by a light signal emitted by the atom to go back to the atom after reflection on the plate; they also
highlight a new transient repulsive character (on the contrary, static electric atom-surface Casimir-Polder forces are attractive). A dynamical Casimir-Polder force has been also considered in the
case of a ground-state atom moving near a flat polarizable surface \cite{DK12}, as well as related dynamical effects due to a fluctuating motion of a wall \cite{BP13}.
All these investigations, which have considered atoms or molecules initially in their bare or partially dressed ground states \cite{MPRSV14}, clearly show that dynamical (time-dependent) aspects yield
new observable features in Casimir-Polder interactions.

In this paper, we investigate the dynamical Casimir-Polder force between an atom and a perfectly conducting plate when the atom is initially in an excited state. This non-equilibrium configuration
shows new relevant features compared to the case of a ground-state atom, due to the presence of an atom-field resonance.
The timescales of the dynamical process (dynamical Casimir-Polder force and dynamical self-dressing) are given by the inverse of the transition frequency of the atom and by the round-trip time.
We use perturbation theory up to second-order in the atom-field coupling, and this limits the validity of our approach to times smaller than the lifetime of the excited atom. We show that the
time-dependent Casimir-Polder force exhibits oscillations in time and changes from attractive to repulsive depending on time and the atom-wall distance. In the asymptotic limit of long times we
correctly recover the known stationary force, given in Refs. \cite{Barton87,Messina2008} for example.
Our results for the atomic excited state give more possibilities to make observable the dynamical effect. First, contrarily to the ground-state case, the static atom-wall Casimir-Polder force on an
excited atom  oscillates in space, yielding spatial regions where the force is attractive and regions where the force is repulsive, while it vanishes at well defined atom-wall distances, whose
position is related to the atomic transition wavelength. Thus, it is possible to single-out more efficiently the dynamical contribution to the force, by choosing specific atom-wall distances where the
static force is zero. Secondly, we show that the dynamical force for the excited atom is much stronger (some orders of magnitude) than the dynamical force for a ground-state atom, using realistic
values of the parameters involved. All these new results thus show that considering an initially excited atom, rather than a bare ground-state atom, should be more a convenient setup for the
experimental detection of the dynamical Casimir-Polder effect.

This paper is organized as follows. In Sec. \ref{sec:level2} we introduce the model and solve the Heisenberg equations for the field and atomic operators.
We then calculate the dynamical Casimir-Polder energy shift for the excited atom up to second-order in perturbation theory as a local atom-field interaction energy, and discuss its asymptotic limit as
well as that of global atomic and field quantities.  In Sec. \ref{sec:level3} we finally obtain the dynamical force and discuss its physical properties and implications. Sec. \ref{sec:level4} is
devoted to our conclusive remarks.

\section{\label{sec:level2} Dynamical energy shift}

Let us consider an atom, modeled as a two-level system, located near a perfectly conducting wall and interacting with the quantum electromagnetic field in its vacuum state. We suppose the atom
prepared in its excited state at $t=0$.
The Hamiltonian describing the atom-field interacting system in the multipolar coupling scheme and within the dipole approximation, is \cite{Passante}
\begin{eqnarray}
\label{eq:1}
&\ & H= H_0+H_I,\\
\label{eq:1a}
&\ & H_0=\hbar\wo S_z+\skj\hbar\wk\adkj\akj,\\
\label{eq:1b}
&\ &H_I = -i\skj\sqrt{\frac{2\pi\hbar c k}{V}}[\bm{\mu}\cdot\fkj(\textbf{r})](S_{+}+S_{-})(\akj-\adkj),
\end{eqnarray}
where $\omega_0=ck_0$ is the atomic transition frequency, $S_z$, $S_{+}$ and $S_{-}$ the pseudospin operators of the atomic system, and $\adkj$ ($\akj$) the creation (annihilation) operators of the
field.
$\bm{\mu}$ is the electric dipole moment of the atom  and $\fkj({\bf r})$ are the field mode functions evaluated at the atomic position. Supposing the wall at $z=0$, the mode functions $\fkj$
taking into account the boundary conditions on a cubic cavity of side $L$ with a wall in the $xy$ plane, are
\begin{widetext}
\begin{eqnarray}\label{eq:2}
(\fkj)_x&=&\sqrt{8}(\ekj)_x\cos\left[k_x\left(x+\frac{L}{2}\right)\right] \sin\left[k_y\left(y+\frac{L}{2}\right)\right]\sin[k_zz], \nonumber \\
(\fkj)_y&=&\sqrt{8}(\ekj)_y\sin\left[k_x\left(x+\frac{L}{2}\right)\right] \cos\left[k_y\left(y+\frac{L}{2}\right)\right]\sin[k_zz], \nonumber \\
(\fkj)_z&=&\sqrt{8}(\ekj)_z\sin\left[k_x\left(x+\frac{L}{2}\right)\right] \sin\left[k_y\left(y+\frac{L}{2}\right)\right]\cos[k_zz],
\end{eqnarray}
\end{widetext}
where $k_x=l\pi/L$, $k_y=m\pi/L$, $k_z=n\pi/L$ with $l,m,n$ positive integers and $\ekj$ are polarization unit vectors. In Eqs. \eqref{eq:2} the cavity walls are located at $x=\pm
L/2$, $y=\pm L/2$, $z=0$ and $z=L$. At the end of the calculation, the case of a single wall at $z=0$ can be recovered by taking the limit $L\rightarrow\infty$.

In order to obtain the time-dependent atom-plate Casimir-Polder force,  we follow a procedure analogous to that in  \cite{VP08,MVP10}.
We first consider the Heisenberg equations for field and atomic operators, and solve them iteratively at the lowest significant order \cite{Power1983}. We obtain
\begin{widetext}
\begin{eqnarray}\label{eq:3}
\akj(t)&=&e^{-i\wk t}\akj (0)+e^{-i\wk t}\sqrt{\frac{2\pi\wk}{\hbar V}}[\bm{\mu}\cdot\fkj(\textbf{r})][S_{+}(0)F(\wo+\wk,t)+S_{-}(0)F(\wk-\wo,t)], \nonumber\\
S_{\pm}(t)&=&e^{\pm i\wo t}S_{\pm}(0)\mp2S_z(0)e^{\pm i\wo t}\skj\sqrt{\frac{2\pi\wk}{\hbar V}}[\bm{\mu}\cdot\fkj(\textbf{r})][\akj (0)F^{*}(\wk\pm\wo,t)-\adkj (0)F(\wk\mp\wo,t)],
\end{eqnarray}
\end{widetext}
where we have defined the function
\begin{equation}\label{eq:4}
F(\omega,t)=\frac{e^{i\omega t}-1}{i\omega}.
\end{equation}

We now calculate the dynamical atom-plate Casimir-Polder interaction during the dynamical self-dressing process of the excited atom.
Using stationary second-order perturbation theory, it is possible to show that the second-order energy shift of the overall system can be obtained as
\begin{equation}\label{eq:5}
\Delta E^{(2)}=\frac 12\bra{\Psi_D}H_I\ket{\Psi_D}
\end{equation}
where $H_I$ is the time-independent interaction Hamiltonian and $\ket{\Psi_D}$ is the dressed (i.e. perturbed) state of the system obtained by second-order perturbation theory.
This relates the energy shift to the local electromagnetic field felt by the atom at its position.
Since we are interested in investigating the time-dependent Casimir-Polder energy shift during the dynamical self-dressing process of the system, we use an appropriate generalization of Eq.
\eqref{eq:5}, that was already used in \cite{VP08,MVP10}.
We first obtain the interaction Hamiltonian in the Heisenberg representation at the second-order in the coupling $H_I^{(2)}(t)$
using Eq. \eqref{eq:3}, and then we evaluate its average value on the initial bare state of the system, that in the present case is the atomic excited state and the photon vacuum
$\ket{\Psi_B} =  \ket{\lbrace0_{\textbf{k}j}\rbrace,\uparrow}$.
Thus, the time-dependent interaction energy is given by
\begin{equation}\label{eq:6}
\Delta E^{(2)}(t)=\frac 12\bra{\Psi_B}H_I^{(2)}(t)\ket{\Psi_B}.
\end{equation}
Our dynamical generalization of \eqref{eq:5} assumes that in our quasistatic approach, the time-dependent interaction energy \eqref{eq:6}, in analogy to the static case, can be
obtained by the local interaction energy between the atom and the field at the specific atomic position. It is worth to note that the overall energy shift of the system is time-independent due to the
unitary time evolution. Nevertheless, as we shall discuss below in more detail, in the limit of large times it gives back the correct expression of the static atom-plate Casimir-Polder force.

Let us now take the atom in
its excited state and the field in the vacuum state as initial configuration at $t=0$, that is $\ket{\Psi_B}=\ket{\lbrace0_{\textbf{k}j}\rbrace,\uparrow}$.
In order to evaluate the local interaction energy \eqref{eq:6}, we first substitute Eqs. \eqref{eq:3} into the interaction Hamiltonian \eqref{eq:1b}; after some algebra, we obtain
\begin{widetext}
\begin{equation}\label{eq:7}\begin{split}
&H_I^{(2)}(t)=-\frac{2\pi ic}{V}\skj k[\bm{\mu}\cdot\fkj(\textbf{r})]^2[S_{+}(0)e^{i\wo t}+h.c]\left\lbrace S_{+}(0)\left[e^{-i\wk t}F(\wo+\wk,t)-e^{i\wk t}F^*(\wk-\wo,t)\right]-h.c\right\rbrace \\
&+\frac{4\pi ic}{V}S_z(0)\skjkj\sqrt{kk'}[\bm{\mu}\cdot\fkj(\textbf{r})][\bm{\mu}\cdot\fkpjp (\textbf{r})]\left\lbrace\akpjp (0)[e^{i\wo t}F^{*}(\wo+\wkp,t)-e^{-i\wo t}F^{*}(\wkp-\wo,t)]+h.c
\right\rbrace \\
&\times [\akj (0)e^{-i\wk t}-h.c].
\end{split}\end{equation}
The dynamical energy shift for the initially excited atom is thus
\begin{equation}\label{eq:8}
\Delta E_{\uparrow}^{(2)}(t)=\frac{\bra{\lbrace 0_{\textbf{k}j}\rbrace,\uparrow}H_I(t)\ket{\lbrace0_{\textbf{k}j}\rbrace,\uparrow}}{2}=-i\frac{\pi c}{V}\skj k
[\bm{\mu}\cdot\fkj(\textbf{r})]^2[e^{-i(\wk-\wo)t}F(\wk-\wo,t)-e^{i(\wk-\wo)t}F^*(\wk-\wo,t)].
\end{equation}
In the continuum limit, after sum over polarizations and angular integrations, we finally obtain
\begin{equation}
\label{eq:9}
\begin{split}
\Delta E_{\uparrow}^{(2)}(d,t)&=-\frac{\mu^2}{4\pi d^3}\int_0^\infty\frac{-2x\cos[x]+(2-x^2)\sin[x]}{x-x_0}\left\lbrace 1-\cos[a(x-x_{0})]\right\rbrace\;dx\\
&=-\frac{\mu^2}{4\pi d^3}\lim_{m\rightarrow 1}\;\left[ \mathcal{D}_m
\left( \int_0^{\infty}\frac{\sin(mx)}{x-x_0} \; dx -
\int_0^{\infty}\frac{\sin(mx)}{x-x_0}\cos[a(x-x_{0})] \;dx \right) \right] .
\end{split}
\end{equation}
\end{widetext}
where $z=d$ is the atom-wall distance,  $x=2kd$, $x_0=2k_0d$, and $a=ct/2d$. Also, we introduced the differential operator $\mathcal{D}_m=(2-2\partial/\partial m+\partial^2/\partial m^2)$. For
simplicity, we have assumed an isotropic atom, i.e $\mu_x^2=\mu_y^2=\mu_z^2=\mu^2/3$.
The contribution coming from the first integral in the second line of Eq. \eqref{eq:9} is time independent, and it coincides with the Casimir-Polder energy shift for an excited atom as obtained with a
time-independent approach. The other integral is time-dependent and it is related to the dynamical dressing of the excited atom. The presence of a pole in $k=k_0$ in the static part
takes into account the possibility of emission of a real photon by the excited atom, and it responsible of the well-known oscillatory behavior of the static atom-wall Casimir-Polder force
\cite{Barton87,Messina2008}. However, in our dynamical calculation there are not poles in the frequency integration, as the first line of \eqref{eq:9} shows.

An explicit evaluation of the time-dependent term in  \eqref{eq:9} shows (see also next Section) that, after a transient yielding a time-dependent Casimir-Polder interaction, it vanishes
asymptotically in time, and the interatomic interaction energy settles to its stationary value. As a result, we correctly recover for $t \to \infty$ the stationary result in
\cite{Barton87,Messina2008}.
We also notice that the energy shift diverges at the round-trip time, that is when $a=2d/ct=1$, similarly to the ground-state case discussed in \cite{VP08}. This behaviour is a manifestation of the
well-known divergences of the radiation reaction and source fields on the light cone in the case of point-like sources \cite{Milonni1976}, and their elimination is still an open problem
\cite{Moniz1977,Milonni94}, as well as the presence of divergences of field energy densities near a reflecting surface \cite{BP12,PRS13,Murray16}. For this reason, in the calculation of the dynamical atom-wall Casimir-Polder force in the next section we will consider separately two different temporal regions: $t<2d/c$
and $t>2d/c$, i.e. before and after the \textit{back-reaction} (or round-trip) time.

There is a conceptually subtle issue about the asymptotic limit $t \to \infty$ that it is worth to stress (it is hereafter understood that our large-times limit in any case considers
times shorter than the decay time of the excited state, of course, due to our perturbative approach). Indeed, the asymptotic approach of expectation values of quantities related to the field and the
atom to the corresponding stationary values should be not taken for granted. Our result in \eqref{eq:9} shows that
\begin{equation}
\label{eq:9a}
\lim_{t \to \infty} \Delta E_{\uparrow}^{(2)}(d,t) = \Delta E_{\uparrow}^{(2)}(d),
\end{equation}
where $\Delta E_{\uparrow}^{(2)}(d)$ is the correspondent stationary value as obtained by the usual time-independent approach. However, a similar relation does not hold for all physical quantities.
For example, we can consider the time-dependent expectation values of the atomic energy $\langle H_A^{(2)}(t)\rangle$, the field energy $\langle H_F^{(2)}(t)\rangle$ and the interaction energy
$\langle H_I^{(2)}(t)\rangle$: their sum is constant due to the unitary time evolution and differs from the static one (bare and dressed excited states have a different energy). Thus the average total
energy of the interacting system does not settle to its static value for $t \to \infty$, as well as the atomic and field parts separately. However, this is not in contradiction with the fact that the
system reaches some \emph{local} equilibrium configuration, as suggested by
\eqref{eq:9a}. In fact, during the self-dressing of the atom, radiation is emitted by the atom \cite{PPP93}; for $t \to \infty$ this emitted field moves to very large distances from the atom and thus
in this limit it does not contribute to the interaction energy with the atom, while contributing to the total field energy. In other words, local field quantities such as its energy density tend to
their stationary values at any finite distance, as confirmed also by calculations of the dynamical Casimir-Polder interaction between two atoms \cite{RPP04}, contrarily to a global quantity such as
the total field energy (that takes also into account the field energy density at a very large distance). Then, we may expect that quantities related to the local field energy-density distribution,
such as the atom-field interaction, should approach the relative stationary value for $t \to \infty$, as confirmed by our result \eqref{eq:9a}.

\section{\label{sec:level3}Dynamical Casimir-Polder Force}

In a quasi-static approach, the dynamical Casimir-Polder force is obtained by taking the derivative of the interaction energy \eqref{eq:9} with respect to the atom-wall distance $d$ and changing its
sign
\begin{equation}\label{eq:10}
F_e(d,t)=-\frac{\partial}{\partial d}\;\Delta E_{\uparrow}^{(2)}(d,t).
\end{equation}
After lengthy algebraic calculation, from \eqref{eq:9} and \eqref{eq:10} we can write the expression of the dynamical Casimir-Polder force as the sum of two terms
\begin{equation}
\label{eq:11}
F_e(d,t)=F_{stat}(d)+F_{dyn}(d,t),
\end{equation}
where $F_{stat}(d)$ is a time-independent contribution coinciding with the asymptotic limit $t \to \infty$ and $F_{dyn}(d,t)$  is the time-dependent part. The first is given by
\begin{widetext}
\begin{eqnarray}
\label{eq:13}
F_{stat}(d) &=& \frac {\mu^2}{12\pi d^4}\left[ 8k_0d-6(2k_0^2d^2-1)\left( f(2k_0d) -\pi \cos (2k_0d) \right) \right. \nonumber \\
&-& \left. 4k_0d (2k_0^2d^2 -3) \left( g(2k_0d) -\pi \sin (2k_0d) \right) \right]
\label{eq:14}
\end{eqnarray}
\end{widetext}
where $f(z)= \text{Ci} (z) \sin (z) -(\text{Si}(z) - \pi /2)\cos (z)$ and $g(z)= -\text{Ci} (z) \cos (z) -(\text{Si}(z) - \pi /2)\sin (z)$ are the auxiliary functions of the sine and cosine integral
functions \cite{AS65}. It coincides with the static force between an excited atom and a conducting wall, as obtained by a time-independent approach \cite{Barton87,MPRSV14}. The
dynamical part is given by
\begin{widetext}
\begin{eqnarray}
\label{eq:15}
F_{dyn}(d,t) &=&
\mu^2  \Big\{ \frac 1{3\pi d^3}\frac {ct\left( 16d^4(9-2d^2k_0^2)+16c^2t^2d^2(-2+d^2k_0^2)+c^4t^4(3-2d^2k_0^2) \right) \sin (ck_0t)}{(4d^2+c^2t^2)^3}
\nonumber \\
&+& \frac 4{3\pi d} \frac {k_0 (-8d^2+c^2t^2) \cos (ck_0t)}{(4d^2-c^2t^2)^2} + \frac 1{12\pi d^4} \big[ [\text{Ci} (2dk_0-ck_0t) + \text{Ci} (2dk_0+ck_0t)] \nonumber \\
&\times& \left( 2dk_0 (3-2d^2k_0^2) \cos (2dk_0) +3(-1+2d^2k_0^2) \sin (2dk_0) \right)
+[\pi + \text{Si}(ck_0t +2dk_0) + \text{Si}(2dk_0-ck_0t)] \nonumber \\
&\times& \left.
\left( 3(1-2d^2k_0^2) \cos (2dk_0) +2dk_0(3-2d^2k_0^2) \sin (2dk_0) \right) \right] \Big\},
\end{eqnarray}
before  the round-trip time ($t<2d/c$), and
\begin{eqnarray}
\label{eq:12}
F_{dyn}(d,t) &=& \mu^2 \Big\{
\frac{-4k_0d^2 \cos (ck_0t)+ct(1-2d^2k_0^2) \sin (ck_0t)}{3\pi d^3(4d^2-c^2t^2)}
-\frac {16dk_0\cos (ck_0t)}{3\pi (4d^2-c^2t^2)^2} \nonumber \\
&+& \frac{ct(64d^4-12c^2d^2t^2 + c^4t^4)\sin (ck_0t)}{3\pi d^3 (4d^2-c^2t^2)^3} + \frac 1{2 \pi d^4} \left[ (\text{Ci}(ck_0t +2dk_0) + \text{Ci}(ck_0t -2dk_0)) \right. \nonumber \\
&\times& \left( 2dk_0 (3-2d^2k_0^2) \cos (2dk_0) +3(-1+2d^2k_0^2) \sin (2dk_0) \right)
+(\text{Si}(ck_0t +2dk_0) - \text{Si}(ck_0t -2dk_0)) \nonumber \\
&\times& \left. \left( 3(1-2d^2k_0^2) \cos (2dk_0) +2dk_0(3-2d^2k_0^2) \sin (2dk_0) \right) \right] \Big\},
\end{eqnarray}
\end{widetext}
for $t>2d/c$, that is after the round-trip time, where $\text{Si}(x)$ and $\text{Ci}(x)$ are respectively the sine integral and cosine integral functions \cite{AS65}.

\begin{figure}[h!]
\centering
\includegraphics[scale=0.42]{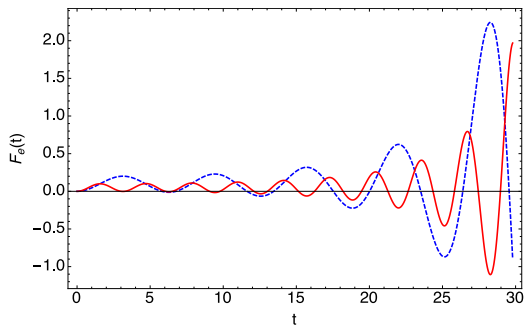}
\caption{(color online) Time evolution of the dynamical Casimir-Polder force on the atom for times smaller than the round-trip time, that is for $t<2d/c$ (force and time are both in arbitrary units). The atom-wall distance is $d=20$, $c=1$, so that the round-trip time is $t=40$. The dashed (blue) and continuous (red) lines represent the force for $k_0=1$ and $k_0=2$, respectively. The plot shows time
oscillations of the force and a strong increase of the force around the round-trip time (where it diverges).}
\label{before}
\end{figure}
\begin{figure}[h!]
\centering
\includegraphics[scale=0.5]{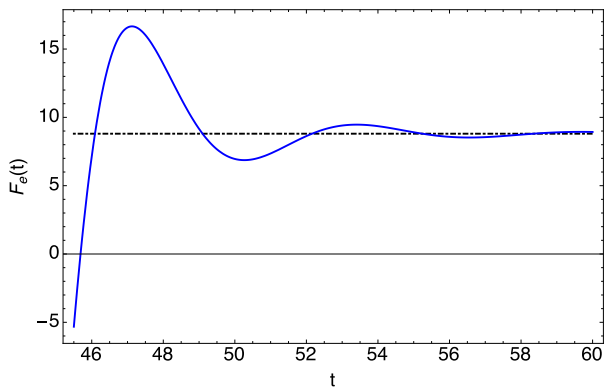}
\caption{(color online) Time evolution of the dynamical force for time $t>2d/c$, that is after the round-trip time (force and time are both in arbitrary units). The atom-wall distance is
$d=20$ and $c=1$, as in Fig. \ref{before}, and $k_0=1$. The time-dependent force shows oscillations around the stationary value represented by the (black) dot-dashed lines. The absolute value of the force strongly increases in the proximity of the round-trip time $t=40$.}
\label{later}
\end{figure}

\begin{figure}[p]
\centering
\subfigure[]{
        \label{Exc_Ground_beforeA}
     \includegraphics[width=0.42\textwidth]{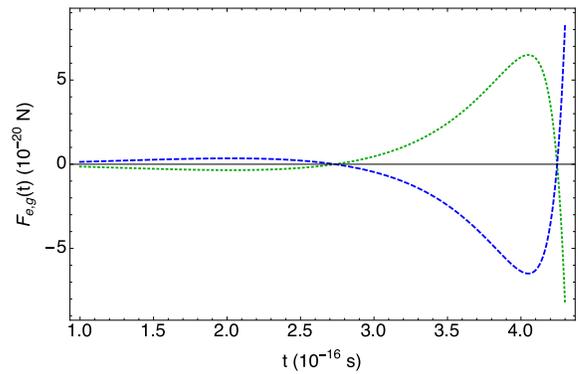} }
\subfigure[]{
        \label{Exc_Ground_beforeB}
        \includegraphics[width=0.42\textwidth]{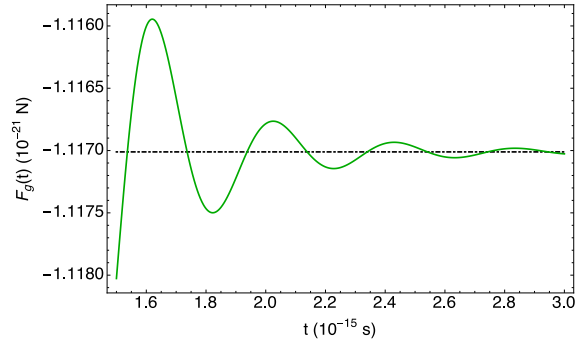} }
\subfigure[]{
        \label{Exc_Ground_beforeC}
        \includegraphics[width=0.42\textwidth]{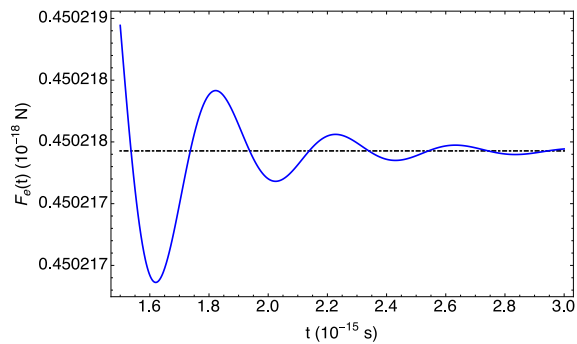}}
\caption{(color online) Comparison of the dynamical Casimir-Polder force between the ground- and excited-atom cases.  Box (a) shows a comparison between the dynamical force for an initially excited
atom (green dotted line) and an ground-state atom (blue dashed line) before the round-trip time $(t<2d/c)$. Boxes (b)and (c) show the dynamical force for the initial bare ground- and excited-state
atom, respectively, after the round-trip time $(t>2d/c)$; in both cases the force for $t \to \infty$ approaches their (nonvanishing) stationary values. In the Figures, the following
typical numerical values for atomic parameters and atom-wall distance have been used: $\mu=6.31\times 10^{-30}$Cm, $\lambda=2\pi/k_0=1.215\times 10^{-7}$m, $d=7.03 \times 10^{-8}$m.
The atom-wall distance has been chosen at the distance where the static force for the excited atom, that is spatially oscillating, reaches its first maximum. Round-trip time is $4.69
\times 10^{-16}$s. Plots in (b) and (c) show that for $t>2d/c$, the dynamical force on the excited atom can be much stronger (three orders of magnitude) than that on the bare
ground-state atom.}
\label{Fig2}
\end{figure}

As expected from our previous physical considerations, the dynamical Casimir-Polder force indeed diverges on the light-cone $t=2d/c$; this is related to the assumptions of a point-like atom (dipole approximation) and a perfectly reflecting mirror, and to our initial condition of a bare state \cite{MVP10}. The divergence would be reduced or smeared out by releasing these assumptions. The dipole approximation and the assumption of a perfect mirror fail at frequencies larger that $c/a_0$, $a_0$ being the Bohr's radius, and the mirror's plasma frequency $\omega_P$, respectively. Such large photon frequencies contribute mainly around the round-trip time $t=2d/c$. We thus expect that our results are not reliable for a time interval around the round-trip time of the order of the largest of $a_0/c$ or $\omega_P^{-1}$ (in order that both assumptions are valid), where the contribution of high-frequency modes is more relevant. For an hydrogen atom and using, for example, the numerical value of the plasma frequency of gold $\omega_P \simeq 1.4 \cdot 10^{16}\, \text{$s^{-1}$}$, we obtain that we cannot safely consider times closer than about $7 \cdot 10^{-17} \,$s to the singularity at $t=2d/c$.

This singular behavior can be easily extrapolated from Figures \ref{before} (for $t<2d/c$) and \ref{later} (for $t>2d/c$).
Also, Fig. \ref{before} shows that the force vanishes at $t=0$ because of our initial condition of a bare (excited) state. In fact, in the case of an initial bare state, the atom-field interaction is suddenly ``switched-on'' at $t=0$; thus, the atomic dynamical dressing, which includes the effect of the conducting wall and yields the dynamical atom-wall interaction, starts at that time too.
If a partially dressed initial state were considered, the force at $t=0$ would not be vanishing, as discussed in \cite{MVP10}.
The fact that the force is not vanishing before the round-trip time (see Fig. \ref{before}) is fully consistent
with relativistic causality because the atom interacts with the vacuum field fluctuations, which are modified by the presence of the reflecting wall from the outset.

The dynamical force shows evident oscillations in time, switching from repulsive (positive) to attractive (negative), as Figures \ref{before} (for $t<2d/c$) and \ref{later} (for $t>2d/c$) show. This
feature is common to the case of an atom initially prepared in a bare or partially dressed ground state near a perfectly conducting wall \cite{VP08,MVP10} or a real surface \cite{HHSRP14}, and for a
chiral molecule \cite{BPRB16}. Figure \ref{later} also shows that in the limit $t\rightarrow\infty$, the known static value of the force for the excited atom as given by a time-independent approach
\cite{Barton87,Messina2008} is recovered. This shows that the internal dynamics of the system, for what concerns with the dynamical Casimir-Polder force and related local field quantities at a finite
distance from the atom and the wall, asymptotically (in time) approaches the equilibrium configuration, as already discussed in Sec. \ref{sec:level2}. The static Casimir-Polder force
between a mirror and an excited barium ion has been observed, and its oscillating behavior has been experimentally confirmed \cite{Wilson03,Bushev04}.

As mentioned, our results are valid for times shorter than the decay time of the excited state, typically of the order of $10^{-8}\, $s. The main effects we have obtained for our dynamical case, i.e. the strong increase of the force around the round-trip time and its oscillations, involve much shorter times, typically of the order of $10^{-16} \,$s and $10^{-15} \,$s, respectively. This shows that the possibility of observing these new effects is fully compatible with the approximations done.

It is worth to compare our results for the excited atom with the results for an initially bare ground-state atom obtained in \cite{VP08,MPRSV14}. Figures \ref{Fig2} show the dynamical force, both
before and after the round-trip time, for the two lowest levels of an hydrogen atom when the atom-wall distance is $d \simeq 7 \times 10^{-8}$m, where the excited-atom static force has
its first maximum. For $t<2d/c$, i.e. before the round-trip time (Fig. \ref{Exc_Ground_beforeA}), the force shows oscillations from attractive to repulsive, that are comparable in
size for the two cases. On the other hand, after the round-trip time, i.e. for $t>2d/c$ (Figures \ref{Exc_Ground_beforeB} and \ref{Exc_Ground_beforeC}),
the intensity of the dynamical force for the excited atom exceeds that for the initially bare ground state by about three orders of magnitude. The physical reason is
that in the excited-atom case, there is, compared with the ground-state case, a much stronger contribution from field modes near resonance with the atomic transition frequency.
Furthermore, the static force for the excited atom is vanishing at specific atom-wall distances (contrarily to the static force for a ground-state atom that is attractive at all distances), and thus
the dynamical term gives the main contribution to the overall atom-wall force around such distances. All this clearly indicates that the case of the excited atom considered in this
paper should be more suitable to probe and detect experimentally the time-dependent dynamical Casimir-Polder effect arising from a non-equilibrium initial state.
Possible experimental setups for observing the dynamical force could be similar to those already used to observe static vacuum level shifts (van der Waals and Casimir-Polder), or changes of the spontaneous decay, for a single trapped atom or ion in the presence of a conducting or dielectric wall \cite{Wilson03,Bushev04,Eschner01,Failache03}. The trapped atom, for example an alkali atom such as Cs or Rb, is initially laser excited using an optical dipole transition, for example the $D_2$ line of Rubidium 87, $5^2D_{1/2} \rightarrow  5^2P_{1/2}$ at $780 \,$nm \cite{Failache03}: the subsequent dynamical Casimir-Polder interaction of the atom with the plate perturbs the harmonic trapping potential and modify the motion of the atom in the trap. Modification of its oscillation frequency in the trap is a signature of its interaction with the wall. A similar method has been also used to verify the temperature dependence of the static Casimir-Polder potential \cite{Obrecht07}.
Finally, we wish to stress that the
intensity of the dynamical Casimir-Polder force could be significantly increased, even by several orders of magnitude, using Rydberg atoms that have high dipole moments (scaling as $n^2$ with the
principal quantum number $n$ \cite{Gallagher88}); also, the space and time oscillations of the dynamical force, being determined by $k_0$, would be much slower in this case, due to the lower
transition frequency, hopefully allowing an easier detection of the transient (repulsive) effects discussed in this paper.

\section{\label{sec:level4}Conclusions}

In this paper we have considered the dynamical (time-dependent) Casimir-Polder force between an atom prepared at $t=0$ in its excited state and a perfectly conducting wall.
We have shown that the dynamical process involves a timescale given by the round-trip time $t=2d/c$ ($d$ being the atom-wall distance), that is the time taken by a light signal emitted by the atom to
reach the atomic position after reflection on the wall. We have evaluated the dynamical force both for times smaller and larger than the round-trip time, and shown that it oscillates in time from
attractive to repulsive. The known static force is recovered for $t \to \infty$ and some subtle questions about the asymptotic approach of dynamical field and atomic quantities to their stationary
values have been discussed. We have also shown that, in the case considered of an initially excited atom, new features appear with respect to the ground-state case already known in the literature,
that could be relevant for the experimental detection of the dynamical effect. In fact, the static Casimir-Polder force for excited atoms vanishes for specific atom-wall distances, where the force
changes its character from attractive to repulsive and vice versa, and thus our dynamical contribution is essentially the main one around such points (this does not occur for a ground-state atom,
because in this case the force is attractive at any distance). Also, our results show that around and after the round-trip time, the dynamical contribution to the Casimir-Polder force is much greater
for the excited atom compared with the known case of a ground-state atom. All these considerations suggest that the dynamical Casimir-Polder force on the excited atom considered in this paper should
be suitable for an easier detection of the dynamical Casimir-Polder effect.

\section*{ACKNOWLEDGMENTS}
The authors wish to thank S. Spagnolo for many discussions on the subject of this paper. FA acknowledges the Marie Curie Actions of the EU's 7$^{\mbox{th}}$ Framework Programme under REA [grant number 317232] for their financial support. RP and LR gratefully acknowledge financial support by the Julian Schwinger Foundation, Universit\`{a} degli Studi di Palermo and MIUR.
SYB and PB were supported by the German Research Foundation (grants BU 1803/3-1 and and GRK 2079/1) and SYB gratefully acknowledges support by the Freiburg
Institute for Advanced Studies.

\end{document}